# ESSMArT Way to Manage User Requests

**Maleknaz Nayebi · Liam Dicke · Ron Ittyipe · Chris Carlson · Guenther Ruhe**

**Abstract** Quality and market acceptance of software products is strongly influenced by responsiveness to user requests. Once a request is received from a customer, decisions need to be made if the request should be escalated to the development team. Once escalated, the ticket must be formulated as a development task and be assigned to a developer. To make the process more efficient and reduce the time between receiving and escalating the user request, we aim to automate of the complete user request management process. We propose a holistic method called ESSMArT. The methods performs text summarization, predicts ticket escalation, creates the title and content of the ticket used by developers, and assigns the ticket to an available developer. We internally evaluated the method by 4,114 user tickets from Brightsquid and their secure health care communication platform Secure-Mail. We also perform an external evaluation on the usefulness of the approach. We found that supervised learning based on context specific data performs best for extractive summarization. For predicting escalation of tickets, Random Forest trained on a combination of conversation and extractive summarization is best with highest precision (of 0.9) and recall (of 0.55). From external evaluation we found that ESSMArT provides suggestions that are 71% aligned with human ones. Applying the prototype implementation to 315 user requests

---

Maleknaz Nayebi
University of Toronto
E-mail: mnayebi@ucalgary.ca

Liam Dicke
University of Alberta
E-mail: dicke@ualberta.ca

Ron Ittyipe
University of Calgary
E-mail: ron.ittyipe@ucalgary.ca

Chris Carlson
Brightsquid Inc.
E-mail: chris.carlson@Brightsquid.com

Guenther Ruhe
University of Calgary - SEDS lab
E-mail: ruhe@ucalgary.ca

resulted in an average time reduction of 9.2 minutes per request. ESSMArT helps to make ticket management faster and with reduced effort for human experts. ES-SMArT can help Brightsquid to (i) minimize the impact of staff turnover and (ii) shorten the cycle from an issue being reported to an assignment to a developer to fix it.

# 1 Introduction

Software evolution is driven by the needs exhibited by its users in the form of feature or maintenance requests. Management of change requests is time consuming and requires significant training and domain experience. Data-driven automation of this process is proposed to increase responsiveness and improve the quality of this data-driven requirements management process [23].

Automated text summarization is the task of producing a concise and fluent summary while preserving key information, content, and overall meaning [1]. A recent survey on the different concepts and techniques was given by Gambhir and Gupta [12]. While initially developed and used outside software engineering, text summarization becomes critical for handling the textual information that is now widely accessible in software development. Automated summarization of bug reports have been studied e.g. by [40]. However, summarization is just the first step in a more comprehensive process of leveraging textual user responses for software product improvement. To increase its practical impact, we increased the scope of automation and propose ESSMArT as a method for automating text summarization, escalation of user requests to the development team, as well as suggesting heading and priority of the automated ticket report.

In this paper, we target the automated escalation, creation, prioritization, and job assignment of development tickets. To do so, we introduce a method called ESSMArT[1]. ESSMArT combines summarization with information retrieval techniques for automated generation of escalated tickets based on user requests. As a case study, we evaluated ESSMArT using data from the development of a Health Communication system offered by a company called BrightSquid. Analyzing customer change requests is an important part of their development process [33]. In particular, from analyzing a set of 4,114 former user requests, we answered the following research questions:

**RQ1: Automated condensing of user requests –** Among existing state-of-the art techniques for condensing user requests by extractive summarization, which one works best in terms of F1 accuracy?

**Why and How:** Typically, once a change request arrives, a Customer Relationship Management (CRM) employee takes over the request and summarizes the request for the user's confirmation. The summary generated is the base for the escalation decision and possibly for creating a development ticket. Automation of this step is intended to reduce the human workload and increase responsiveness using the state of the art ROUGE metric for evaluation.

---

[1]ESSMArT: **ES**calation and **SuM**marization **AuT**omation

**RQ2: Predicting escalation of user requests —** Comparing three classification algorithms Naive Bayes, Support Vector Machines and Random Forest, which one works best for predicting the escalation of user requests?

**Why and How:** Support for Customer Relationship Management (CRM) staff in predicting escalation is expected to help in terms of (i) effort needed and (ii) quality of prediction. For the prediction, we compared three machine learners that have proven successful in similar contexts.

**RQ3: Quality of automatically generated ticket content —** How well are the ESSMArT generated ticket titles and contents aligned with the ones generated by human experts?

**Why and How:** In practice, often the CRM manager is the only person who creates development tickets after deciding on their escalation. These tickets are more general than summarizing the conversation and are represented by a title and a body that describes the problem or the requested enhancement. ESSMArT uses abstractive summarization for creating the ticket title and uses a thesaurus to generate development tickets from the summary of tickets studied in **RQ1**.

**RQ4: Quality of operationalization —** In comparison to the results of a human expert, how correct are (i) the predicted priorities of the tickets and (ii) the assignments of the tickets generated by ESSMArT to developers?

**Why and How:** The priority of a ticket decides how urgent the ticket should be handled. If the priority is assigned correctly, the tickets will be handled in the correct order. The proper assignment of the ticket to an available developer is critical to get the change request implemented. This currently is a manual process depending on the expertise of the developers. We benchmarked with three states of the art classifiers to automate the process by using the analogy of the upcoming ticket with the former tickets.

**RQ5: Usefulness of ESSMArT for experts —** Utilizing a prototype implementation of ESSMArT, how much are the generated results aligned with the perception of the human experts and how useful are the results?

**Why and How:** External evaluation looks into the perceived valued of ESSMArT from using a prototype implementation. A set of 315 actual user requests were executed by 21 CRM experts and 33 project managers. We measured execution time and surveyed the perceived usefulness of the tool.

The paper is subdivided into eight sections. In Section 2, we first provide more details on the context and motivation for this research. Related work is analyzed in Section 3. The ESSMArT method is outlined in Section 4. The results for different types of (internal) validation of the method is the content of Section 5. In addition, external validation was pursued with results given in Section 6. Subsequently, in Section 7, we discuss limitations and threats to validity of the research. We conclude and give an outlook to future research in Section 8.

## 2 Context and Motivation: User Request Management at Brightsquid

Brightsquid Secure Communication Corp[2] is a global provider of HIPAA-compliant[3] communication solutions - providing compliant messaging and large file transfer for medical and dental professionals since 2009. Secure-Mail is Brightsquid's core communication and collaboration platform. It is offering role-based API access to a catalog of services and automated workflows. It supports aggregating, generating, and sharing protected health information across communities of healthcare patients, practitioners, and organizations. Brightsquid has been working on a number of projects in this directions and this study is focused on analyzing four of these systems. The company is facing the typical problem of software start-ups: The need to quickly enter a competitive market with innovative product ideas while having the goal of short-term revenue generation by satisfying current users and their expectations. At the same time, the company is facing the demand of enlarging their customer base [17].

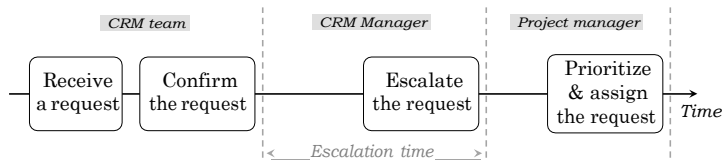

Fig. 1: Change request management process at Brightsquid

The BrightSquid process of managing change requests is shown in Figure 1. When a new user request arrives, a CRM staff member decides if the request should be transferred into a development ticket or not. If yes, the CRM manager escalates the customer request by summarizing the user request and translating it into technical language. As the result of escalation, a new ticket is opened for the software development team. Then, the project manager defines the issue type and adds the ticket to the backlog of customer requests. The stack of user requests is a set of Jira tickets tagged as "CRM escalated" within the backlog. In each bi-weekly sprint at Brightsquid, the project manager scans through all the escalated tickets, discusses the technical aspects of the ticket, and decides whether or not to assign the ticket to the development team. If the ticket is not assigned then the request is maintained in the backlog of tasks to be solved at a later date.

Between *November,* 2014 and *June,* 2017, Brightsquid recorded 4,114 user requests. 7.8% of these users requests were escalated to the development team. These change requests constitute 10.7% of the whole backlog (including 3,026 tickets overall) over these 32 months. Mining the time stamp data of their ticket system, we identified that the escalation by the CRM manager is the process bottleneck. The time that a change requests wait for CRM manager to decide on the request's escalation (escalation time) is on average 26.6% of the total time from receive to assigning the ticket. Moreover, the escalation process needs the involvement of the CRM manager and the project manager [35].

## 3 Related Work

Motivated by the problem at Brightsquid, we propose ESSMArT for managing user requests. The scope of this investigation is defined by the time the customer request arrives and includes escalating the request as well as finally assigning it to a developer to solve the problem. To the best of our knowledge, this is the first study that analyses the process as a whole and across different repositories involved in the process. Former studies have focused exclusively on predicting if a ticket is escalated, or only on summarizing the bug reports. Below, we provide an overview of the existing works.

### 3.1 Escalating user requests

Bruckhaus et al. [6] performed one of the very first studies in ticket escalation by providing an early model of predicting the risk of a ticket escalation in *Sun Microsystems*. The study intended to reduce the risk and cost of the development process by not escalating user reported defects that are known in the system. As the follow up of this study authors took a business-oriented perspective in [22] and created a prediction system to predict escalations of known defects in enterprise software to maximize the return on investment (ROI). They limited their study to defect escalation with the intent of preventing risky and costly escalations. The study used a decision tree classifier to find the most cost-effective method, which had a significantly higher ROI than any other method. Also focused on the cost of escalating defects, Sheng et al. [43] used a cost-sensitive decision tree algorithm to obtain the lowest possible cost in dealing with defects. This paper, similarly to the previous one includes negatives in the cost matrix to account for the benefits of correct classifications.

More recently, Montgomery and Damian [29] performed a study with IBM analyzing ticket escalation. The study is focused on selecting the right attributes to predict ticket escalation accurately. They used data such as forming a detailed customer profile to help determine if the customer was likely to escalate a given ticket, including many details like the customer's expected response time compared to the average response time of the analysts. Using a set of 2.5 million support tickets they were able to achieve a recall of 0.79. The model that they used in order to predict accurately at such a scale involved focusing more on what features to use rather than using all available data. In addition, Montgomery et al. [30] suggested the tool ECrits to mitigate the information overload in the decision process of escalating user request. Managing user requests for mobile applications was the content of a survey performed by Martin et al. [26]. However, the notion of escalation was not included as these studies are just focusing on analyzing and prioritizing user requests.

### 3.2 Text summarization

Automated text summarization methods are usually discussed under two general categories of *extractive*, and *abstractive* text summarization. For a recent survey of these two categories, see the work of Das and Marins [8].

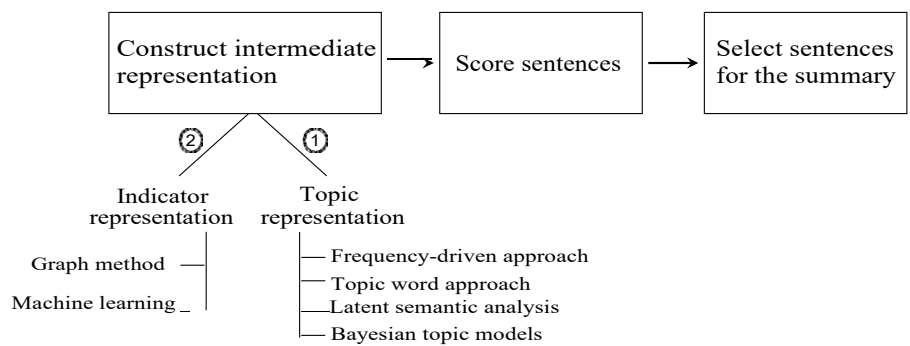

Fig. 2: Overview of extractive summarization methods inferred from [1]. Summarization methods mainly differ in the way they construct the intermediate representation.

### 3.2.1 Extractive text summarization

Extractive text summarization refers to a method of taking a pre-existing document and extracting the sentences that best make up the content of the document. These sentences are taken word for word from the original document. This process is guided by a variety of factors such as the frequency of the words in the sentence, or the similarity of the sentence to the title of the article [13].

Extractive summarization has three major steps. First, an intermediate representation of the text is constructed. Second, the sentences are scored based on their calculated importance. In the third step, the sentence score is used for ranking sentences, and the sentences with the highest rank would be selected to be part of the extractive summary. Extractive summarization techniques use multiple features and different feature weights for selecting the most representative sentences for the summary. An overview of the extractive summarization process and methods are illustrated in Figure 2. Extractive summarization methods are different in terms of constructing the intermediate representation. Two major representation techniques exist: *topic representation* and *indicator representation*.

Nazar et al. [36] provided an overview of the literature of software artifacts. In software engineering mostly extractive methods of summarization have been used. Text summarization is most often applied to large documents such as news reports [5] to allow for concise reading of the entire document. Researchers have been benchmarking and adopting summarization techniques to improve the accu- racy of summarizing software artifacts. In Table 1 we provide an overview of the most related papers to ours, their application domains, their used summarization techniques, and the size of their dataset.

### 3.2.2 Abstractive text summarization

Abstractive text summarization refers to the summarization of text passages and documents utilizing one of many corpus-backed NLP methods. The ultimate goal is to synthesize sentences based on sentence generation, which is done after clus-tering, importance determination, and other information extraction techniques or ranking methods running on top of an underlying language model. Abstractive NLP summarization techniques most commonly utilize a large corpus and subsequently generated language model, allowing for abstractive methods to perform information extraction and ranking [44].

## 4 ESSMArT for Automated User Request Management

The design of the method was inspired by real-world projects at Brightsquid. ESSMArT consists of five main steps. While some of these steps (e.g., summarization, escalation) were adapted from existing work, the main value of our method is that it provides a holistic approach covering the whole process starting from the arrival of a change request up to issuing a ticket and assigning it to a developer. As part of the method development, we studied and compared different variants of implementing these steps. The five steps of ESSMArT are shown in Figure 3. In the below subsections, we describe these main steps in more detail.

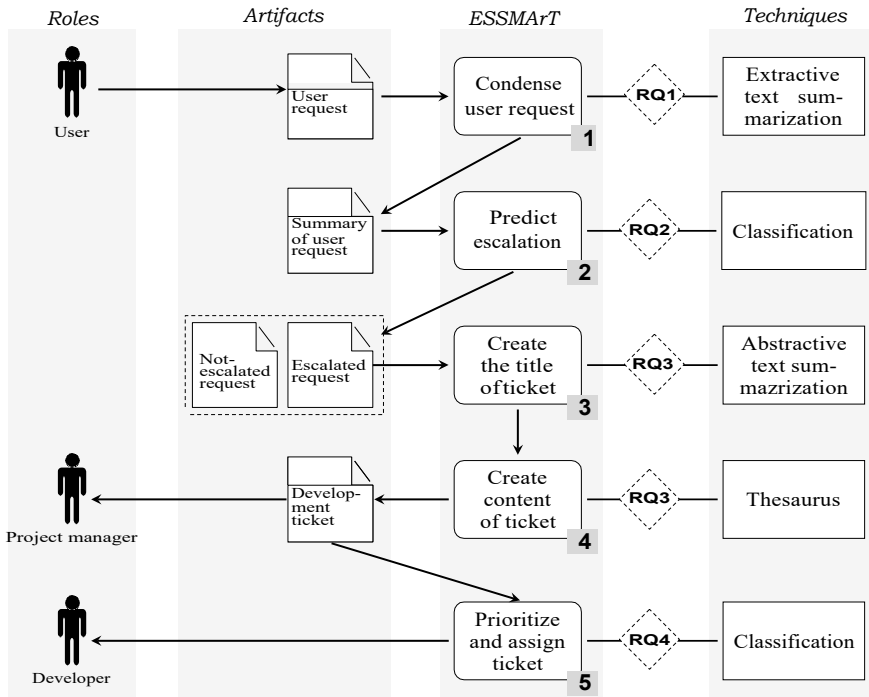

Fig. 3: Process of ESSMArT for automated generation of developer tickets from user requests

### 4.1 Condense user requests

Whenever a request comes from a customer, first the CRM staff who receives the request summarizes the incoming request and then checks with the customer if the summarization correctly reflects the request of the customer. Thus, in Step 1 of ESSMArT in Figure 3 we start with automated summarization of the users request.

Looking into the 4,114 user requests, we found that they were initially submitted to the CRM and included on average 9.3 (median = 8.8) sentences. Similarly, the summaries created by the CRM staff included on average 4.7 sentences (median = 4.1). Furthermore, we compared the length of tickets generated by eight different CRM staff members and could not find any significant difference from running the ANOVA test ($P - value = 0.31$). This means that the length of the tickets is about half of the length of the original conversation and is independent of the person in the CRM team who receives and responds to the customer request. To condense the conversation in the form of a user ticket we used extractive summarization. As the result of this analysis, we limited the number of sentences for our extractive summarization to five sentences.

To select the best summarization method in Step 1 of ESSMArT, we compared various extractive summarization methods. The methods were selected based on their popularity in literature, in particular in software engineering research. In addition, we looked at the availability of the summarization methods as open source tools or libraries. For the frequency driven approach we used the SumBasic algorithm suggested by Vanderwende et al. [46]. On the topic word approach, we applied Edmundson method [11]. For Latent Semantic Analysis (LSA) we applied Steinberger and Jezek method [45]. For Bayesian topic modeling, we used *Textrank* suggested by Mihalcea and Tarau [27]. The summary of the used methods and the pointer into the related literature is presented in Table 1.

Also similar to Rastkar et al. [40], we used machine learning to build an indicator representation. We built classifiers using Naive Bayes [41] as it proved to outperform other classifiers for summarization tasks [1]. We trained the classifier only on the content of the tickets and based on three datasets (i) Enron dataset of emails as it includes the conversation between human subjects [7], [32]. One of the authors annotated the summary of this dataset, (ii) The conversation between developers and users of Ubuntu on Fedora channel (denoted as UDC) with similar nature as the conversations at Brightsquid, and (iii) the conversations between CRM team and customers of Brightsquid (denoted as BSC).

Table 1: Related software engineering research that used extractive summarization.

| Application domain | Paper | Summarization technique | Dataset |
|---|---|---|---|
| Bug reports | Rastkar et al. [39] [40] | Machine learning | 36 Mozilla Bug report |
| Bug reports | Mani et al. [24] | Graph method | 55 DB2 bug reports |
| App reviews | Di Sorbo et al. [9] | Topic classification | 17 mobile apps |
| Release notes | Moreno et al. [31] | Pre-defined heuristic | 1,000 release notes |

To guide among the existing summarization techniques, we performed a pairwise comparison of the extractive summarization methods. Second, we compared the results of automated extractive summarization with the summaries created by human experts. For this purpose, we used Recall-Oriented Understudy for Gisting Evaluation (ROUGE) [21]. ROUGE is the most widely used method for evaluation of the summarization quality. While ROUGE has different variations, we followed [1] and used *ROUGE-n* and *ROUGE-SU*.

*ROUGE-n* is based on the comparison of n-grams. Within each comparison, one of the summaries is considered as the reference and the other summary, also known as the candidate, is compared against it. Within this comparison process, ROUGE-n elicits bi-grams and tri-grams:

$$ROUGE\_n = \frac{P}{q} \tag{1}$$

Where p represents the number of common n-grams between the two candidate summaries and q represents the number of n-grams that were extracted from reference summary only. *ROUGE-SU* elicits both bi-grams and uni-grams and allows the insertion of words in bi-grams. In other words, the bi-grams do not need to be consecutive sequences of words.

## 4.2 Predict escalation

In Step 2 of ESSMArT, we predict if a ticket should be escalated or not (See Figure 3). To this end, we trained and compared the performance of three classifiers. The three classifiers used were Support Vector Machine [14], Naïve Bayes [41], and Random Forest [20]. All three techniques have been successfully used before in software classifications. Here, they were used to classify tickets as escalating or non-escalating. For training the classifiers, we used Scikit's package of Python. When applicable, we applied an exhaustive search over classifier parameters (such as *Kernel*, *Gamma*, and *C* values) in a way to maximize the score of the left out data. We used the GridSearchCV (GridSearchCV optimizes parameters by cross-validated grid-search) function of Scikit and leveraged both textual content of the requests as well as non-textual content:

(i) *Using textual content of inquiries:* We trained and compared classifiers using the term frequency-inverse document frequency (TF-IDF) values of the words that make up the tickets. TF-IDF is a statistical measure frequently used in information retrieval and text mining to evaluate the importance of words in a collection of documents [38]. It consists of two components, Term Frequency (TF), which is a count of the number of times a word appears in a document, normalized by the total number of words in that document. The second component is the Inverse Document Frequency (IDF) which is the logarithm of the number of documents in the corpus divided by the number of documents where the particular term appears.

Table 2: Extractive summarization techniques analyzed for Step 1 of ESSMArT.

| ID | Class | Type | Description |
|---|---|---|---|
| SumBasic | Topic repres. | Freq. driven | Proposed by Vanderwende et al. [46] used in [16], [48] |
| Edmundson | Topic repres. | Topic driven | ——- |
| Steinberger | Topic repres. | Latent semantic | ——- |
| LDA | Topic repres. | Bayesian models | ——- |
| TextRank | Indicator repres. | Graph method | ——- |
| Enron | Indicator repres. | Machine learning | Trained a Naive Bayes classifer on the Enron email dataset[4] |
| UDC | Indicator repres. | Machine learning | Trained a Naive Bayes classifer on the Ubuntu-related conversations on the Freenode IRC network[5] |
| BSC | Indicator repres. | Machine learning | Trained a Naive Bayes classifer on the Brightsquid CRM conversation with customers |

Table 3: Attributes associated with customers' request

| ID | Name | Description |
|----|------|-------------|
| $Att_1$ | Conversation | Information provided in the ticket's conversation |
| $Att_2$ | Requester | The individual who requested the ticket originally |
| $Att_3$ | Ticket type | What type of request was made |
| $Att_4$ | Tags | What tags are present in the ticket |
| $Att_5$ | Via | What medium the ticket was introduced through |
| $Att_6$ | Severity | The extent to which the ticket affects the product |
| $Att_7$ | Assignee | Who was assigned to handle the ticket |
| $Att_8$ | Time open | How long it has been since opening a ticket |
| $Att_9$ | Time escalated | How long it has been since escalating the ticket |
| $Att_{10}$ | Time to assign | How long the ticket takes to be assigned |
| $Att_{11}$ | Subject | Content contained within the subject |
| $Att_{12}$ | Brand name | Which product the ticket relates to |
| $Att_{13}$ | Organization | Which organization is requesting the ticket |

We compared classifiers based on different text attributes. We compared the TF-TDF with (i) Bag of Words (BOW) [47] as a representation of the conversations, (ii) the extractive summaries, and (iii) a combination of the two. All the three options were studied using both lemmatized and non-lemmatized tickets. All classifiers were run using the TF-IDF of the conversation and the extractive summaries.

(ii) *Using non-textual attributes for prediction:* We elaborated on the textual content of the users' inquiries by using other attributes recorded alongside it in predicting the escalation and priority of tickets. To evaluate if using any of the recorded data, such as requester, organization, and the time stamp (see the full list in Table 3), could increase the accuracy of the classifiers or not we used Minimum Redundancy Maximum Relevance (mRMR) algorithm [37]. mRMR is an algorithm to select features for classifiers with the intention of minimizing attributes' redundancy. We used R implementation mRMR algorithm for this purpose. mRMR is superior to methods such as information gain analysis [50] as mRMR[6] considers the relation between the attributes as well.

---

[6] Comparing mRMR with Principle Component Analysis (PCA) [49] and Independent Component Analysis (ICA) [15], mRMR does not need the mapping of features into the orthogonal and independent space.

| | | |
|---|---|---|
| Original Sentence | | John emailed me and wanted a copy of a message note faxed to him. |
| Transformation Steps | *In the EMR system;*  *a doctor*   *a doctor*  *specifying system* | |
| Transformed Sentence | | In the EMR system; a doctor emailed a doctor and wanted a copy of a message note faxed to him. |

Fig. 4: An example of transforming a sentence from user request to a development ticket using ESSMArT.

### 4.3 Create a ticket title

Each development ticket consists of a title and a body that describes the problem. In Step 3 of ESSMArT, we used abstractive summarization to suggest ticket titles. The knowledge base used for abstractive NLP summarization techniques most commonly utilize a large corpus and subsequently generated language model that allows for abstractive methods to perform information extraction and ranking. We implemented the abstractive summarization using AbTextSumm [4]. This method was designed and initially implemented by Banerjee et. al [4].

### 4.4 Create ticket content

In Step 4 of ESSMArT, we built a thesaurus and used that to map the user language into the terms understandable by the developers. In a request, users report their experience on using the software. However, a development ticket reflects the high level story in a way that is more understandable for the development team (for example to reproduce bugs). When a customer communicates with a CRM team, only the name and surname of the customer needs to be specified and the CRM team will refer to their customer base to get the related info such as the role of the user, organization, and brand name[7]. When creating the development ticket the content should be self explanatory. To make this happen we take the summary of the conversation that we created in Step 1 of ESSMArT and:

(i) Specify brand name: The development ticket should explicitly mention which product the ticket relates to. As a result, each ticket starts with "in the XYZ systems" within which, XYZ is the name of the system. The user, associated with the request, allows on which systems to have access to.

(ii) Specify the role of the requester: The development ticket should specify the role or the requester to reflect the story. We found this by mining the satellite data around each user. For example, if "Jane" is calling from "Crowfoot clinic" we find in the organization chart that she is the "administrator".

(iii) Abstract specific names to general entities: The specific names should be replaced and mapped to a known general entity (instead of "John Doe", it should be "patient"). In the case that we fail to map a specific name into a general entity, we eliminate it from the sentence.

Figure 4 shows this transition for a sample sentence. To make these mappings we built a thesaurus based on the Brightsquid data. To build this thesaurus we specified a set of documents that contain the related information about our enti-    ties.
We used the *user stories*, *release notes* from Brightsquid as well as descriptions of
the organizations using Brightsquid products. We used word co-occurrence and
grammatical dependencies [42] using the Stanford NLP toolkit [25]. Then we de-
tected the specific names within the summaries using rule-based Named-Entity
Recognition (NER) [28]. Rule-based NER uses a set of named entity extraction
rules for different types of named entity classes with an engine that applies the
rules and the lexicons to the text [28].

---

[7]We focused on the four products of Brightsquid among all the developed systems by them

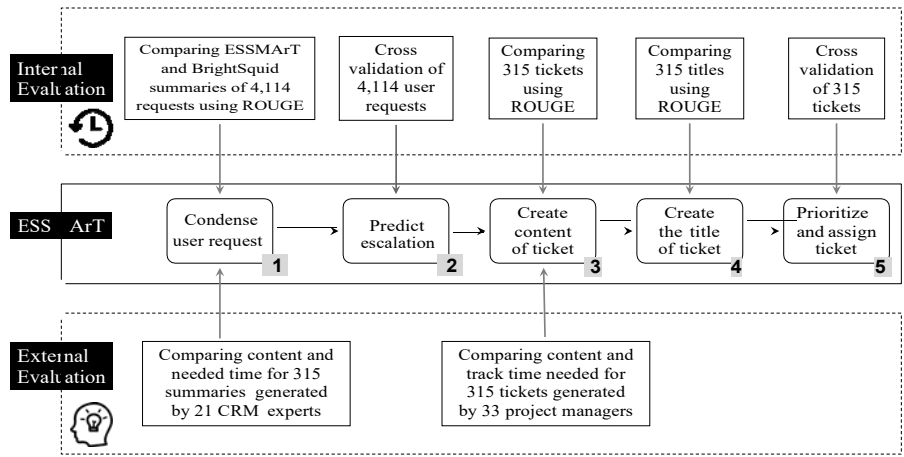

Fig. 5: Evaluation of ESSMArT

## 4.5 Prioritize and assign tickets

Once the ticket is created, each should be assigned a priority in the backlog of
the project, the options being Blocker, Critical, Major, Minor, or Trivial. Then, a
developer should be assigned to the ticket to solve and close it. Prioritization and
assignment of the tickets are well established problems in software engineering
[18], [3]. An assistant for creating bug report assignment recommendations was
proposed by [2]. It was shown that quite reliable bug recommendations can be
offered even with limited project knowledge.

in Step **5** of ESSMArT, for both assignment and prioritization of the ticket,
we reasoned by analogy. We built and compared state of the art classifiers (Naive
Bayes, SVM, and Random Forest) to predict the priority of the ticket and assign
it to a developer. To find the analogy between the ticket and formerly assigned
and prioritized tickets, we used mRMR method to select attributes among the list of
Table 3 to train the classifiers. The details have been discussed in the method for
Step **2** (See Section 4.2).

## 5 Internal evaluation

The proposed method has been evaluated at the different steps and from different perspectives (i.e., internal and external perspective). An overview of all evaluation done is given in Figure 5. In this section, we discuss the different aspects of internal evaluation of ESSMArT. The analysis refers to RQ1, RQ2, RQ3, and RQ4. By its nature, this all is a retrospective analysis based on Brightsquid data.

### 5.1 RQ1: Summarization of user requests

To condense the conversation in the form of a user ticket we used extractive summarization. We applied and compared the results of eight summarization techniques. These eight techniques covered different extractive summarization classes based on unsupervised (topic representation, indicator representation) and supervised learning (see Table 2). For the supervised learning, we trained a Naive Bayes classifier on three different datasets (Email communications, Dialogues between developers and users of Ubuntu, and Brightsquid's conversation with the users). We compared the eight techniques pairwise based on ROUGE and ROUGE-SU measures. Table 4 shows the result of pairwise comparison of the summarization techniques. Within this table, each row is compared with the technique in the column being considered as the baseline summary. As a general trend, the results of supervised learning methods (Enron, UDC, and BSC) are performing closely similar to each other comparing to unsupervised methods. Among the unsupervised methods, TextRank almost consistently worked better than the others.

We further compared these methods by comparing these eight different summarization methods with summaries from CRM experts at Brightsquid. As illustrated in Figure 6 we found that the supervised extractive summarization trained on Brightsquid data is performing the best but it is only 4.2% better in terms of F-score than the classifier trained with the Ubuntu dataset (UDC). TextRank as an unsupervised learning method performs as the third best method in our case study. Overall, Steinberger is the worst performing classifier. It is 20% less accu- rate than BSC in terms of the F-score. Considering the effort needed to prepare

Table 4: F1-score of extractive techniques for summarizing user requests.

| Method | ROUGE | | | | | | | | ROUGE-SU | | | | | | | |
|---|---|---|---|---|---|---|---|---|---|---|---|---|---|---|---|---|
| | SumBasic | Edmondson | Steinberger | LDA | TextRank | Enron | UDC | BSC | SumBasic | Edmondson | Steinberger | LDA | TextRank | Enron | UDC | BSC |
| SumBasic | – | 0.78 | 0.8 | 0.8 | 0.82 | 0.76 | 0.73 | 0.71 | – | 0.74 | 0.76 | 0.77 | 0.79 | 0.7 | 0.66 | 0.63 |
| Edmondson | 0.79 | – | 0.76 | 0.86 | 0.83 | 0.77 | 0.73 | 0.73 | 0.75 | – | 0.72 | 0.82 | 0.8 | 0.73 | 0.69 | 0.64 |
| Steinberger | 0.78 | 0.85 | – | 0.86 | 0.79 | 0.74 | 0.7 | 0.66 | 0.74 | 0.79 | – | 0.82 | 0.77 | 0.68 | 0.68 | 0.65 |
| LDA | 0.82 | 0.78 | 0.81 | – | 0.8 | 0.79 | 0.75 | 0.7 | 0.79 | 0.74 | 0.77 | – | 0.78 | 0.7 | 0.66 | 0.6 |
| TextRank | 0.79 | 0.73 | 0.78 | 0.8 | – | 0.74 | 0.69 | 0.67 | 0.77 | 0.7 | 0.75 | 0.78 | – | 0.69 | 0.65 | 0.6 |
| Enron | 0.65 | 0.66 | 0.7 | 0.7 | 0.73 | – | 0.81 | 0.79 | 0.69 | 0.7 | 0.71 | 0.7 | 0.71 | – | 0.83 | 0.81 |
| UDC | 0.64 | 0.63 | 0.68 | 0.65 | 0.63 | 0.94 | – | 0.91 | 0.61 | 0.61 | 0.63 | 0.6 | 0.61 | 0.85 | – | 0.84 |
| BSC | 0.6 | 0.61 | 0.6 | 0.59 | 0.61 | 0.95 | 0.96 | – | 0.55 | 0.55 | 0.52 | 0.56 | 0.59 | 0.88 | 0.88 | – |

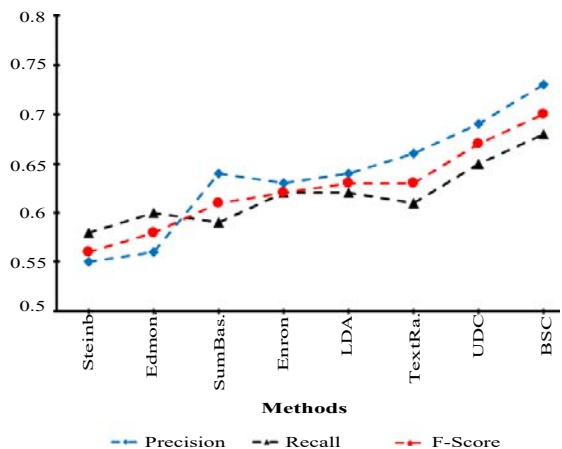

Fig. 6: Performance of extractive summarization techniques in comparison with human generated summaries using ROUGE-SU

training sets, unsupervised methods are proven to be an alternative to supervised methods.

> *Supervised learning based on context specific Brightsquid data performs best for summarizing user request and outperforms the best unsupervised technique by 10%.*

## 5.2 RQ2: Predicting escalation

We compared three state of the art classifiers that exhibit good performance with a short text to predict the escalation of a user's requests, the priority of the tickets, and the developers assigned to it.

We used mRMR to select non-textual content for predicting escalation. However, we did not find any of the $Att_2$ to $Att_{13}$ significantly increasing the accuracy of the predictive model (Step 2 of ESMMArT). This is aligned with the current practice at Brightsquid. In Brightsquid the content of the conversation, specifically the communicated concern by the user, is the only decisive factor for the CRM team to escalate a ticket to the development team. As the result, we focused our effort on automating escalation only on the content of the tickets.

To deal with the imbalanced number of escalated and non-escalated tickets, we down-sampled the non-escalated conversations. We benchmarked different techniques for increasing the accuracy of the textual similarity analysis such as customizing the list of stop words, lemmatization, stemming, BOW (bag of words), TF-IDF, and using extractive summaries for predicting escalation. Overall, we found that by using tf-idf the F1-score of the three classifiers is better by 8.7% on average. Also, we found that by using lemmatization instead of porter's stemming the F1-score were improved by 4.3% on average. Our results also showed that eliminating a customized set of stop words from the conversation increases the accuracy of the classifier by 6.1% on average. Table 5 summarizes the results of the classification techniques using the 4,114 conversations to predict ticket escalation. Within the tables, the numbers in italic font represent the corresponding top values.

Table 5: Accuracy of classification algorithms for predicting tickets' escalation.

| **Naive Bayes** | Precision | Recall | F1 |
|---|---|---|---|
| Conversation | 0.83 | 0.45 | 0.58 |
| Conversation + Lemmatization | 0.85 | 0.49 | 0.62 |
| Extractive summary + Lemmatization | 0.82 | 0.43 | 0.57 |
| Conversation + Extractive summary + Lemmatization | 0.86 | 0.53 | 0.65 |
| **Support Vector Machine (SVM)** | Precision | Recall | F1 |
| Conversation | 0.64 | 0.43 | 0.51 |
| Conversation + Lemmatization | 0.62 | 0.45 | 0.52 |
| Extractive summary + Lemmatization | 0.60 | 0.42 | 0.49 |
| Conversation + Extractive summary + Lemmatization | 0.62 | 0.49 | 0.54 |
| **RandomForest** | Precision | Recall | F1 |
| Conversation | 0.88 | 0.49 | 0.62 |
| Conversation + Lemmatization | 0.89 | 0.53 | 0.66 |
| Extractive summary + Lemmatization | 0.89 | 0.42 | 0.56 |
| Conversation + Extractive summary + Lemmatization | 0.90 | 0.55 | 0.68 |

The classifiers used for classification were Support Vector Machine (SVM), Naive Bayes, and Random Forest. For each of them, four alternatives were evaluated. Looking at the F1 measure as the one balancing precision and recall, we found that the combination of just looking at the conversation and the extractive summary in combination with lemmatization performs most promising (two times best out of three). When comparing the classification techniques, Random Forest performs best in terms of precision, and best in F1 on the best configuration. In contrast, SVM seems to be the lowest performing among the three techniques overall.

> *Random Forest classifier trained on a combination of conversation and extractive summarization outperforms the other models in terms of precision. Using the extractive summaries increases the F-score of the prediction.*

### 5.3 RQ3: Quality of automatically generated ticket content

Each development ticket consists of content and a title. ESSMArT suggests the content using a thesaurus for mapping and generalizing entities having extractive summaries of **RQ1** as input and using abstractive summarization to suggest titles. We report the results of evaluating the content and title of these tickets.

*Evaluating the content of the development tickets:* To bridge between the customer request and the development ticket we built a thesaurus that maps the user terminology and specific names into the developers' terminology or general entities.

For building the thesaurus we used 304 separate documents including user stories, release notes, organization and brand descriptions, and team descriptions. As the result of this automated process, we built a thesaurus with 3,301 entries. We manually went through this thesaurus and in particular searched for the personnel names. We mapped the specific personnel name into the organizational roles. For each distinct specific name (1,908 out of 2,467) we entered three separate entries for the first name, surname and the name as a whole. We ended up with a thesaurus with 7,117 entries in total. We then used ROUGE-SU for evaluating the alignment of the generated tickets with the tickets extracted by human experts.

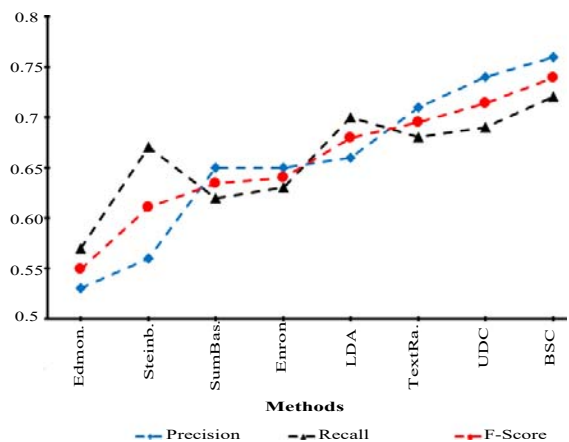

Fig. 7: Comparison of the content of the development tickets generated by human experts with the ones generated with different summarization techniques and the thesaurus in RQ3 using ROUGE-SU.

Figure 7 shows the results of the comparison between automated and manually created tickets. In a majority of cases, the precision of the classifiers is better than their recall. Looking into the F-Score, the combination of the supervised learning method and thesaurus performs best. Interestingly, unsupervised methods based on the graph model of represented indicators (TextRank [27]) combined with the use of our thesaurus performed well.

> *Supervised learning trained on Brightsquid data combined with the use of a thesaurus performs best among all techniques. The accuracy is 5% better than the best unsupervised method in terms of F1-score.*

*Evaluating the title of the development tickets:* We compared the titles created using abstractive summarization with the human-generated titles using ROUGE-SU. Within the process of creating the ticket titles, we limited the size of the abstractive summary to 11 words as it was the average length of the ticket titles in Brightsquid.

Table 6: Ticket titles using ESSMArT summarization in comparison with human experts using ROUGE-SU.

| Abstractive summarization | Precision | Recall | F1 |
|---|---|---|---|
| Enron | 0.61 | 0.57 | 0.59 |
| UDC | 0.56 | 0.53 | 0.54 |
| BSC | 0.68 | 0.63 | 0.65 |

We presented the results of this comparison in Table 6. Training the model based on the Brightsquid data performed the best among the other models having an F-score of 0.65.

## 5.4 RQ4: Prioritization and assignment of tickets

We compared three state of the art classifiers with good performance proven in other contexts. We used short texts to predict the priority of the escalated tickets and assign it to a developer.

### 5.4.1 Prediction of the priority of escalated tickets

Several attributes are recorded along with the user requests as shown in Table 3. The mRMR analysis showed the importance of *organization* and *brand name* for predicting the priority of the tickets. We compared three classifiers, Naive Bayes, Support Vector Machine (SVM), and Random Forest. We benchmarked the performance of these classifiers using the conversation between users and Brightsquid, the extractive and abstractive summaries of the apps, as well as the organization and brand name attributes. Similar to the escalation prediction in the previous section, we evaluated the impact of different text pre-processing and processing methods. In all cases, lemmatization was applied. The results of our benchmark are shown in Table 7.

When using the textual content of the conversation only, Random Forest classifiers have a slightly better performance in comparison to Naive Bayes. Using extractive and abstractive summaries for predicting the tickets priority increased the F1-score of the classifiers up to 8.6% on average. Having abstractive summarization on top of that further increased the F1-score. Moreover, using *organization* and *brand name* increased the F1-score by up to 12.3%. When comparing across all pre-processing and processing options, Naive Bayes with using both textual and non-textual features performs best.

Table 7: Accuracy of classification algorithms for predicting tickets' priority.

| **Naive Bayes** | Precision | Recall | F1 |
|---|---|---|---|
| Conversation | 0.64 | 0.61 | 0.62 |
| Extractive summary | 0.64 | 0.62 | 0.63 |
| Conversation + Extractive summary | 0.68 | 0.63 | 0.65 |
| Abstractive summary + Extractive summary | 0.68 | 0.66 | 0.67 |
| Conversation + Abstractive summary + Extractive summary | 0.68 | 0.66 | 0.67 |
| Abstractive summary + Extractive summary + Organization + Brand name | *0.74* | *0.72* | *0.73* |
| **Support Vector Machine** | Precision | Recall | F1 |
| Conversation | 0.52 | 0.51 | 0.51 |
| Extractive summary | 0.52 | 0.53 | 0.52 |
| Conversation + Extractive summary | 0.55 | 0.53 | 0.54 |
| Abstractive summary + Extractive summary | 0.56 | 0.53 | 0.54 |
| Conversation + Abstractive summary + Extractive summary | 0.58 | 0.55 | 0.56 |
| Abstractive summary + Extractive summary + Organization + Brand name | 0.70 | 0.67 | 0.68 |
| **Random Forest** | Precision | Recall | F1 |
| Conversation | 0.64 | 0.62 | 0.63 |
| Extractive summary | 0.66 | 0.63 | 0.64 |
| Conversation + Extractive summary | 0.66 | 0.63 | 0.64 |
| Abstractive summary + Extractive summary | 0.69 | 0.66 | 0.67 |
| Conversation + Abstractive summary + Extractive summary | 0.70 | 0.68 | 0.69 |
| Abstractive summary + Extractive summary + Organization + Brand name | 0.73 | 0.70 | 0.71 |

### 5.4.2 Assignment of tickets to developers

Similar to the what we did for predicting prioritization, we compared the three state of the art classifiers with multiple textual attributes. The results of this benchmarking are shown in Table 8. The results showed that similar to the ticket prioritization, using the content of the conversation along with the abstractive and extractive summaries performs the best and Naive Bayes outperforms SVM and Naive Bayes in this task.

We also used mRMR to select none textual features (as listed in Table 3 and found *brand name* as an important factor for ticket assignment. On average, using the *brand name* on top of the textual features increased the F-Score of the classifiers.

> *Using Naive Bayes perform the best for prioritizing (F-score = 0.73) and assigning the tickets (F-score = 0.86). The results showed that using extractive and abstractive summarization along with other features increases the accuracy of these predictions.*

Table 8: Accuracy of classification algorithms for assigning tickets to developers.

| **Naive Bayes** | Precision | Recall | F1 |
|---|---|---|---|
| Conversation | 0.8 | 0.83 | 0.81 |
| Extractive summary | 0.81 | 0.83 | 0.82 |
| Conversation + Extractive summary | 0.85 | 0.83 | 0.84 |
| Abstractive summary + Extractive summary | 0.85 | 0.84 | 0.84 |
| Conversation + Abstractive summary + Extractive summary | 0.87 | 0.84 | 0.85 |
| Abstractive summary + Extractive summary + Brand name | *0.89* | *0.84* | *0.86* |
| **Support Vector Machine** | Precision | Recall | F1 |
| Conversation | 0.68 | 0.61 | 0.64 |
| Extractive summary | 0.68 | 0.62 | 0.65 |
| Conversation + Extractive summary | 0.68 | 0.63 | 0.65 |
| Abstractive summary + Extractive summary | 0.69 | 0.65 | 0.67 |
| Conversation + Abstractive summary + Extractive summary | 0.71 | 0.67 | 0.69 |
| Abstractive summary + Extractive summary + Brand name | 0.74 | 0.68 | 0.71 |
| **Random Forest** | Precision | Recall | F1 |
| Conversation | 0.72 | 0.73 | 0.72 |
| Extractive summary | 0.76 | 0.73 | 0.74 |
| Conversation + Extractive summary | 0.76 | 0.75 | 0.75 |
| Abstractive summary + Extractive summary | 0.78 | 0.74 | 0.76 |
| Conversation + Abstractive summary + Extractive summary | 0.8 | 0.78 | 0.79 |
| Abstractive summary + Extractive summary + Brand name | 0.83 | 0.78 | 0.8 |

## 6 External evaluation

So far, we built and compared the state-of-the-art techniques known for the different stages in the process of managing user requests. As a form of retrospective analysis, we called that *internal evaluation*. However, the question of the perceived value of applying the method is still open. In this section, ESSMArT is evaluated by CRM experts and project managers. The section is closely related to RQ5, and is called *external evaluation*. The subjects are asked whether ESSMArT makes the process of escalation faster and better. As we did not have access to employees of Brightsquid, we recruited 21 CRM experts and 33 project managers from outside. We used convenient sampling for recruiting participants from social media to participate in this study. The whole external evaluation is described in the subsequent subsections.

### 6.1 Protocol for external evaluation of ESSMArT

To evaluate the performance of ESSMArT, we asked four CRM experts and two project managers to go through the escalation of a sample set of user requests, first without and then with using ESSMArT. Offering the task through social platforms, we attracted 21 CRM experts and 33 project managers to participate. We assigned 15 escalated user requests to each CRM expert to perform the evaluation.

For the evaluation, we first performed a manual process for escalating tickets and then we used ESSMArT:

*Manual process:* We provided the complete conversation of 315 Brightsquid user requests and asked to provide a summary. Each CRM expert (i) evaluated 15 anonymized conversations and (ii) was asked to decide if the ticket should be escalated. Furthermore, we asked her to provide an extractive summary by se-lecting a subset of sentences of the conversation, without applying any rewording [40]. We recorded the time needed per ticket called $Escalation_{time}$.

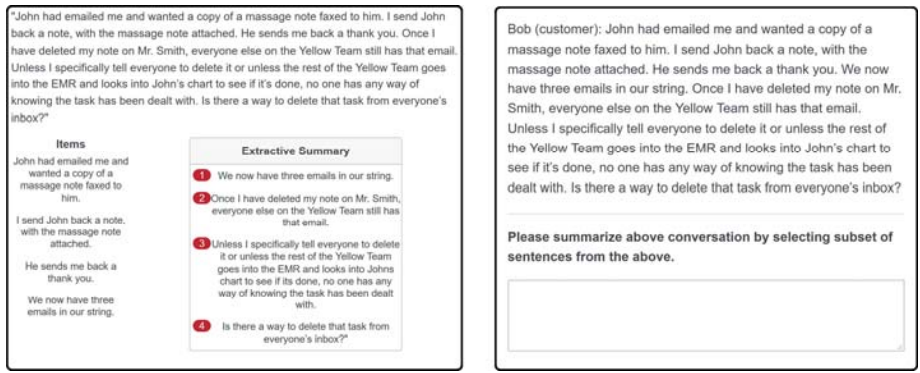

Fig. 8: Process of manual escalation for evaluation of ESSMArT. The left screen shot was shown to the CRM experts while the right one was shown to the project managers.

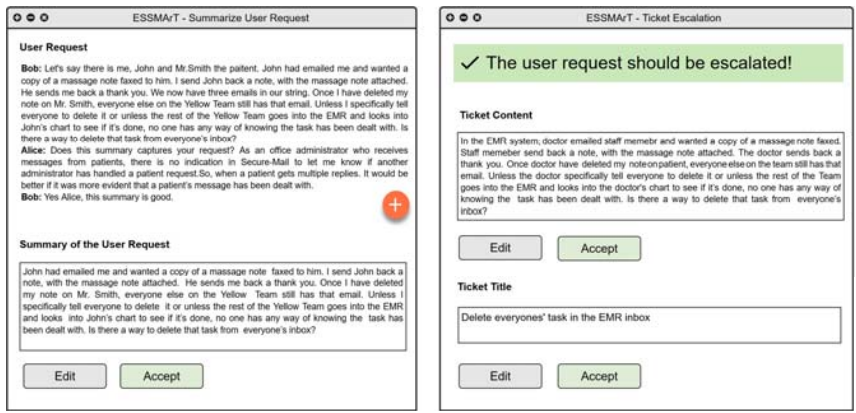

Fig. 9: Applying the prototype tool of ESSMArT, the left figure shows the summarization of a sample user request. The right one suggests the content and title of the sample development ticket.

We submitted the summary of 315 user requests to the 33 project managers participating in our experiment. Each project manager prepared a development ticket based on the summarized user request. We recorded this time as $Decision_{time}$. The screenshot of the survey for manual escalation is shown in Figure 8.

*ESSMArT way:* We provided the prototype implementation of ESSMArT to the CRM experts and the project managers. We logged the time that it has taken them to perform each step. To make the results comparable with the manual process, we did not allow any CRM expert or project manager to work on any ticket she already has handled in the manual process. Figure 9 shows the screenshot of ESSMArT for a sample request. The left screen shows the ESSMArT UI for CRM experts used to log the $Escalation_{time}$ while the right one is the UI shown to the project managers and logged as $Decision_{time}$.

## 6.2 RQ5: Usefulness of ESSMArT for Experts

For the external evaluation, we surveyed experts to find the usefulness of the results and to figure out if the process would be faster for humans using our prototype tool.

### 6.2.1 Are the ESSMArT results aligned with perception of the external experts?

We compared the sentences selected by CRM experts with those selected by ESSMArT. Based on the results of internal evaluation in Section 5, we used Random Forest trained with former BrightSquid data for summarization. We asked survey participants in the role of CRM to select most representative sentences. Comparing the selected sentences selected by ESSMArT and CRM experts resulted in 0.71 precision and 0.77 recall ($F1 - score = 0.73$). Figure 10 - (a) shows the distribution of the number of sentences that were selected differently between ESSMArT and human experts.

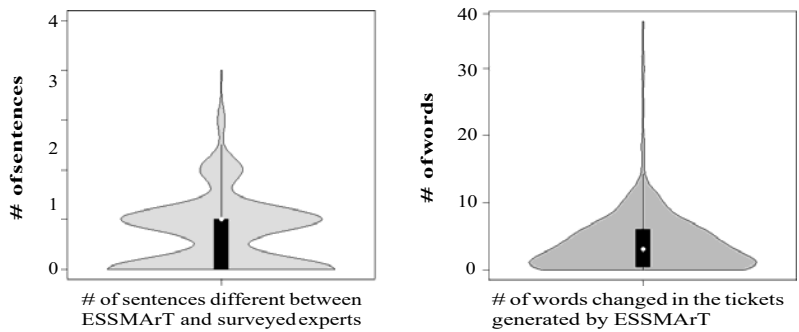

Fig. 10: Analysis of the response to the survey questions from 21 CRM experts and 33 project managers in terms of the number of different sentences (left) and the number of words changed (right).

To evaluate the alignment of the tickets generates by ESSMArT with those generated by human experts, we tracked the number of words changed by project managers on the ESSMArT generated ticket. Figure 10 - (b) shows the distribution of the changed words for the 315 user requests. In 25% of the tickets no word has been changed and in 8.5% of the cases, more than 10 words were changed. Overall, across 315 user requests, 3.7 words on average were changed by human experts.

Figure 11 shows the results of the questions we asked survey participants. We asked the 54 survey participant to state if they agree that the ESSMArT results were understandable. Only 1.9% (one) of the participants disagreed with this statement. 51.9% of the participants stated they likely or very likely would use ESSMArT in practice. Trusting decision support tools is a common problem in their usability [10]. 68.6% of participants stated that they trust the ESSMArT results while 7.5% of them stated it is unlikely that they would trust the results.

> *ESSMArT provides suggestions that are 71% aligned to the selection of human experts. The average change of 7.5 words per ticket also demonstrates the usefulness of its results for project managers.*

### 6.3 Does ESSMArT make the escalation process faster?

We logged the $Escalation_{time}$ and $Decision_{time}$ for 315 user requests when done completely manual and by using ESSMArT. Figure 12 shows the the time taken for each of these tasks. Figure 12 - (a) shows the $Escalation_{time}$ for CRM experts with and without ESSMArT. Using ESSMArT reduces $Escalation_{time}$ by 3.2 minutes on average, per ticket:

(i) $ESSM\ ArT_{(Escalation_{time})} < Manual_{(Escalation_{time})}$: For 297 (94.2%) of the requests, ESSMArT allowed CRM experts do the task faster.

(ii) $ESSM\ ArT_{(Escalation_{time})} \geq Manual_{(Escalation_{time})}$: For 18 of the user requests ESSMArT appeared not helpful in making the process faster. In these cases, the manual process took on average 0.35 seconds less time for escalation. Considering the small number of cases and small difference and the ticket have been escalated by two different participants for manual and ESSMArT enabled process, we believe these cases happen because of difference between the cognitive ability of the participants and possible distractions.

Similarly, we logged and compared the $Decision_{time}$ for the participated project managers. In this case, the $Decision_{time}$ has been improved in all cases. Using ESSMArT allowed project managers to decide on average 6.3 minutes faster in comparison to the manual process. Figure 12-(b) shows the boxplot for the time taken by project managers to decide on a ticket with and without using ESSMArT.

Independently, we asked the survey participants how likely ESSMArT usage would reduce the cycle time of change requests. 48.2% of the participants agreed or strongly agreed that ESSMArT reduces the time needed for escalating a user request. Of the CRM experts, 54.4% agreed and 30.3% strongly agreed that ESSMArT make their job faster and 42.9% of the project managers agreed with 23.8% strongly agreeing about the same for their escalation tasks. Figure 12-(c) shows the total time saved for each expert across all the tickets.

> *ESSMArT reduces the escalation time of a change request by 9.2 minutes on average. 84.7% of the CRM experts and 66.7% of project managers agreed or strongly agreed that ESSMArT helps them to perform the task faster.*

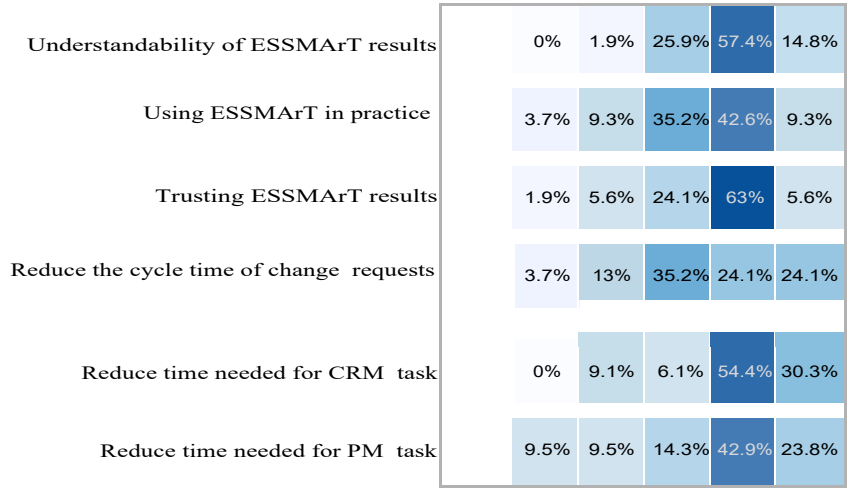

Fig. 11: Analysis of responses from 21 CRM experts and 33 project managers having participated in the survey.

# 7 Limitations and threats to validity

The ESSMArT process described in Figure 2 is general enough to be applicable beyond BrightSquid. One key prerequisite for the applicability is the existence of both a user ticket as well as a developer ticket system. In any case, the results of our study should be treated with caution due to the existing threats to validity discussed below.

*Construct validity* - To measure the quality of summarization techniques for condensing user request we used the ROUGE metrics as it was shown valid results in similar contexts before. In most of the comparisons, we compared the result of summarization with the content generated by human experts at Brightsquid, retrospectively. In former studies, with the intent to find the best summarization method, researchers asked external developers to take the bug reports and select a subset of the sentences. We extended the scope of the evaluation and intended to provide evidence that the results of our automation could be used to assist human experts.

*External validity* - Selection of techniques for ESSMArT and validation of the different steps were closely connected with the real-world data set of 4,114 tickets. At each step of ESSMArT we compared and selected among state of the art methods. However, our study and the results are limited to the context of BrightSquid. The challenge for testing ESSMArT in other contexts is access to the holistic dataset of the whole process. This limits our ability to provide evidence on the generalizability of the method. However, to the best of our knowledge, the process of managing customer requests is not unique to BrigthSquid which makes ESSMArT useful for other software companies.

Also, we used convenient sampling which imposes the risk of a selection bias and thus causing a lack of credibility in general. However, it was considered acceptable as it just served as an initial evaluation for exploratory purposes [19].

*Internal validity* - To provide evidence that ESSMArT makes the escalation process faster, we used experts from outside Brightsquid. We mitigated this risk by careful screening of the participants and their background.

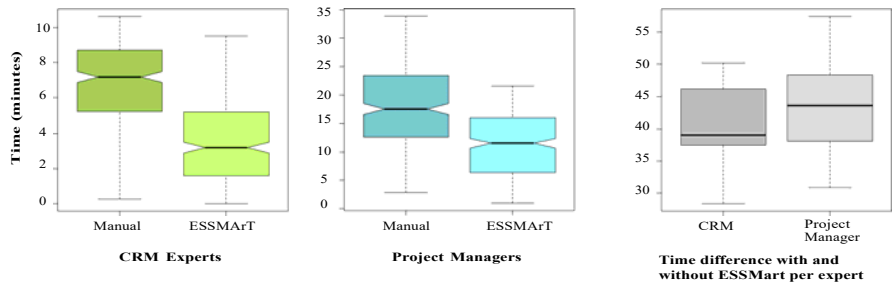

Fig. 12: The logged time when the escalation is done manually versus escalation using ESSMArT. This is the sum of the time taken by CRM experts and by project managers in categorizing 315 change requests.

*Conclusion validity* - We analyzed the tickets of Brightsquid within a limited time frame available. The methods of the company might change over the lifetime of the projects and the customers. While we compared the length of the tickets for different CRM employees, the CRM manager, and the project manager were always the same person. A change in the staff might slightly change the results presented in this paper caused by the difference in their interaction with customers, text, and tickets. Also, For a few user requests (18 cases) ESSMArT appeared not helpful in making the process faster. Considering the small number of cases and small we believe these rare cases happen because of the difference between the cognitive ability of the participants and possible distractions.

## 8 Conclusions and Future Work

ESSMArt addresses user request management from a holistic perspective and supports decision-makers by providing them with intelligent suggestions within the process. We looked into the whole process from the time receiving a request or question from a user to the time that the problem is resolved by the CRM or a task is assigned to a developer. This is unique as the existing research has been focused on the steps of this process in isolation only. The method development was inspired by the industrial collaboration project with BrightSquid, However, its underlying process is following the general steps of user request management and thus is applicable more broadly. We believe that automating the management of user requests and their escalation would increase the chance of innovation within organizations [34].

We consider the results as necessary, but not sufficient for claiming external validity. More empirical evaluation of the individual steps of the method as well as on the impact of the whole method is required. In particular, as the data is coming from one company only, the evaluation needs to be extended to other environments. Also, the existing prototype tool was intended to perform an initial evaluation and needs to be further enhanced and more comprehensively tested.

## Acknowledgment

This research was partially supported by the Natural Sciences and Engineering Research Council of Canada, NSERC Discovery Grant RGPIN-2017-03948 and the NSERC Collaborative Research and development project NSERC-CRD-486636-2015. We appreciate the discussion with and suggestions made by *Lloyd Montgomery* to make the paper better.